%% file: macs0647jd.tex
\documentclass[twocolumn]{aastex631}

\usepackage{}
\usepackage{gensymb}  
\usepackage{multirow}  

\shorttitle{Carbon Abundance of MACS0647--JD}
\shortauthors{Hsiao et al.}

\begin{document}

\title{First direct carbon abundance measured at $z>10$ in the lensed galaxy MACS0647$-$JD}

\correspondingauthor{Tiger Hsiao}
\email{tiger.hsiao@cfa.harvard.edu}

\newcommand{\CfA}{\affiliation{Center for Astrophysics \text{\textbar} Harvard \& Smithsonian, 60 Garden Street, Cambridge, MA 02138, USA}}

\newcommand{\STScI}{\affiliation{Space Telescope Science Institute (STScI), 3700 San Martin Drive, Baltimore, MD 21218, USA}}

\newcommand{\JHU}{\affiliation{Center for Astrophysical Sciences, Department of Physics and Astronomy, The Johns Hopkins University, 3400 N Charles St. Baltimore, MD 21218, USA}}

\newcommand{\ESAAURA}{\affiliation{Association of Universities for Research in Astronomy (AURA), Inc.~for the European Space Agency (ESA)}}

\newcommand{\Austin}{\affiliation{Department of Astronomy, University of Texas at Austin, 2515 Speedway, Austin, Texas 78712, USA}}

\newcommand{\BGU}{\affiliation{Physics Department, Ben-Gurion University of the Negev, P.O. Box 653, Be'er-Sheva 84105, Israel}}

\newcommand{\kapteyn}{\affiliation{Kapteyn Astronomical Institute, University of Groningen, 9700 AV Groningen, The Netherlands}}

\newcommand{\CAB}{\affiliation{Centro de Astrobiología (CAB), CSIC-INTA, Ctra. de Ajalvir km 4, Torrejón de Ardoz, E-28850, Madrid, Spain}}

\newcommand{\Cambridge}{\affiliation{Kavli Institute for Cosmology, University of Cambridge, Madingley Road, Cambridge CB3 0HA, UK}}

\newcommand{\Cavendish}{\affiliation{Cavendish Laboratory, University of Cambridge, 19 JJ Thomson Avenue, Cambridge CB3 0HE, UK}}

\newcommand{\UCL}{\affiliation{Department of Physics and Astronomy, University College London, Gower Street, London WC1E 6BT, UK}}

\newcommand{\Arizona}{\affiliation{Department of Astronomy / Steward Observatory, University of Arizona, 933 N Cherry Ave, Tucson, AZ 85721}}


\author[0000-0003-4512-8705]{Tiger Yu-Yang Hsiao} \CfA \JHU \STScI 

\author[0000-0001-8426-1141]{Michael W. Topping} \Arizona

\author[0000-0001-7410-7669]{Dan Coe} \STScI \ESAAURA \JHU

\author[0000-0002-0302-2577]{John Chisholm} \Austin

\author[0000-0002-4153-053X]{Danielle A. Berg}\Austin

\author[0000-0002-5258-8761]{Abdurro'uf} \JHU \STScI

\author[0000-0002-7093-1877]{Javier Álvarez-Márquez} \CAB

\author[0000-0002-4985-3819]{Roberto Maiolino} \Cambridge \Cavendish \UCL

\author[0000-0001-8460-1564]{Pratika Dayal} \kapteyn

\author[0000-0001-6278-032X]{Lukas J. Furtak} \BGU





\input{newcommands}


\begin{abstract}
Investigating the metal enrichment in the early universe helps us constrain theories about the first stars and study the ages of galaxies.
The lensed galaxy MACS0647$-$JD at $z=10.17$ is the brightest galaxy known at $z > 10$.
Previous work analyzing JWST NIRSpec and MIRI data yielded
a direct metallicity $\rm{12+log(O/H)}=7.79\pm0.09$ ($\sim$ 0.13 $Z_\odot$)
and electron density $\rm{log}(n_e / \rm{cm^{-3}}) = 2.9 \pm 0.5$,
the most distant such measurements to date.
Here we estimate the direct C/O abundance for the first time at $z > 10$,
finding a sub-solar ${\rm log(C/O)}=-0.44^{+0.06}_{-0.07}$.
This is higher than other $z>6$ galaxies with direct C/O measurements, 
likely due to higher metallicity.
It is also slightly higher than galaxies in the local universe with similar metallicity.
This may suggest a very efficient and rapid burst of star formation, a low effective oxygen abundance yield, or the presence of unusual stellar populations including supermassive stars.
Alternatively, the strong \CIIIdw\ emission ($14\pm 3\,{\rm \AA}$ rest-frame EW)
may originate from just one of the two component star clusters JDB ($r \sim 20$ pc).
Future NIRSpec IFU spectroscopic observations 
of MACS0647$-$JD will be promising for disentangling C/O in the two components
to constrain the chemistry of individual star clusters just 460 Myr after the Big Bang.

\end{abstract}


\section{Introduction} \label{sec:intro}
The first generation of stars (Pop {\sc iii}) are believed to contain no elements heavier than helium \citep[dubbed metals; e.g.,][]{Barkana2001,Klessen2023}.
The quest to find Pop {\sc iii} stars is ongoing and will help us understand how the first metals were built up.
Since directly detecting Pop {\sc iii} stars is challenging, understanding the chemical abundance of heavy elements in high-redshift galaxies could be key to constraining the properties of Pop {\sc iii} stars.
Specifically, oxygen and carbon are the most abundant metals in the universe.
Oxygen is generated heavily from the death of massive stars ($M>8\,M_{\odot}$) through core-collapse supernovae (CCSN), soon after the onset of star formation \citep[e.g.,][]{Nomoto2013}.
Carbon is enriched not only via CCSN but also during the asymptotic giant branch (AGB) phase of intermediate-mass stars which {have longer} lifetimes than massive stars that undergo CCSN \citep[e.g.,][]{Kobayashi2011,Karakas2014,Kobayashi2020}.
Therefore, the C/O ratio can be diagnostic of galaxy ages, increasing at ages $\gtrsim$ 100$\,$Myr.
Other astrophysical processes also affect C/O.
For instance, the C/O ratio is sensitive to 
supernova-driven outflows \citep{Berg2019},
star-fomration history (SFH) \citep{Berg2020},
and the initial mass function (IMF).
High-z galaxies are thought to have a top-heavy initial mass function (IMF) \citep{Inayoshi2022}, favoring the formation of more massive stars and a lower C/O.
Meanwhile, the oxygen abundance, O/H, traces the integrated star formation history.
Comparisons of the C/O and O/H abundances constrain the relatively recent star formation history (through C/O) versus the total star formation in the galaxy (O/H).


The JWST, a groundbreaking space telescope, opens the window to spectroscopically study the chemical abundances in high redshift galaxies in high redshift galaxies ($z>6$) thanks to its unique wavelength coverage, unparalleled sensitivity, and spatial resolution in the rest-frame optical \citep[e.g.,][]{Arellano2022,Jones2023,Isobe2023,DEugenio2024,Bunker2023,Maiolino2023,Cameron2023,Hsiao2024a,Topping2024,Castellano2024,Hsiao2024b}.
Most of these galaxies have low metallicity ($Z<0.2\,Z_{\odot}$) and low C/O ratios (${\rm log(C/O)<-0.6}$) as expected for low-mass galaxies that are heavily impacted by stellar feedback from relatively recent star formation \citep{Arellano2022,Jones2023,Isobe2023,Stiavelli2023,Topping2024,Curti2024,Marques2024,Topping2025}.
Surprisingly, {a high C/O ratio of ${\rm log(C/O)>-0.17}$} was discovered in galaxy GS-z12 at $z=12.5$, and was interpreted as the heritage of Pop {\sc iii} stars \citep{DEugenio2024}, since supermassive stars (e.g. Pop {\sc iii} stars) could have different abundance patterns than the typical massive stars.
Regarding other $z>10$ galaxies, GHZ2 ($z=12.34$) has a subsolar carbon abundance of ${\rm log(C/O)\sim-0.94}$ to ${-0.53}$ \citep{Castellano2024}, and log(C/O)$>-0.78$ is estimated in GN-z11 \citep[$z=10.6$;][]{Cameron2023}.
However, all previous C/O measurements at $z>10$ were derived from assumed electron temperature ($T_{e}$) and/or assumed electron density ($n_{e}$) due to the lack of resolved temperature-sensitive and density-sensitive line ratios.
To reduce the uncertainty, direct C/O measurements are needed to truly understand how carbon is enriched. 

MACS0647$-$JD, a lensed galaxy at $z=10.17$, provides a promising laboratory for the C/O abundance in the early universe.
It is triply lensed to three images: JD1, JD2, and JD3, with magnifications of 8, 5.3, and 2.2, respectively.
Given its high magnification (F200W AB mag 25.0), \citet{Hsiao2023a} studied JWST NIRCam imaging, which resolved MACS0647$-$JD into two small components JDA and JDB, suspected to be a possible galaxy merger, plus a possible third companion C (\about 3$\,$kpc away).
JDA is larger (delensed radius of $r=70\pm24\,{\rm pc}$), brighter, and bluer, likely due to its young stellar population ($\sim50\,{\rm Myr}$ old) and lack of dust.
In contrast, JDB, is smaller (delensed effective radius of $r = 20^{+8}_{-5}\,{\rm pc}$), and is redder, likely due to an older stellar population ($\sim$100 Myr old) and mild dust ($A_{V}\sim0.1\,{\rm mag}$).
An additional possible companion JDC is nearby, 3$\,$kpc away.
Later, \citet{Hsiao2024a} and \citet{Hsiao2024b} reported JWST Cycle 1 NIRSpec prism spectroscopy and JWST Cycle 2 MIRI IFU spectroscopy.
The auroral line \OIIIwa\ and \OIIIw\ were detected in NIRSpec prism and MIRI IFU spectroscopy, respectively, yielding the first direct metallicity measurement at $z>10$ of $12+{\rm log(O/H)}=7.79\pm0.09$ \citep{Hsiao2024b}.
Here, we combine the \CIIIdw\ detections from NIRSpec prism data with the \OIIIwa\ and \OIIIw\ observations from NIRSpec and MIRI to provide the first direct C/O measurement at $z>10$. 
These data explore the enrichment of stellar populations within galaxies in the first few hundred Myr of cosmic time.

Throughout this article, we adopt solar abundance ratios
\logOH\ = 8.69 and ${\rm log(C/O)}=-0.23$ \citep{Asplund2021}.
Lensing magnifications of 8.0 and 5.3 are adopted for JD1 and JD2, respectively \citep{Hsiao2023a}.
Magnification uncertainties (\about 15\%) do not affect line flux ratios or derived abundance ratios.
Where needed, we adopt the {\em Planck} 2018 flat \LCDM\ cosmology \citep{Planck18_cosmo}
with $H_0 = 67.7$ km s\inv\ Mpc\inv, $\Om = 0.31$, and $\OL = 0.69$,
for which the universe is 13.8 billion years old
and $z = 10.17$ is 460 Myr after the Big Bang.


\begin{figure*}
\centering
\includegraphics[width=\textwidth]{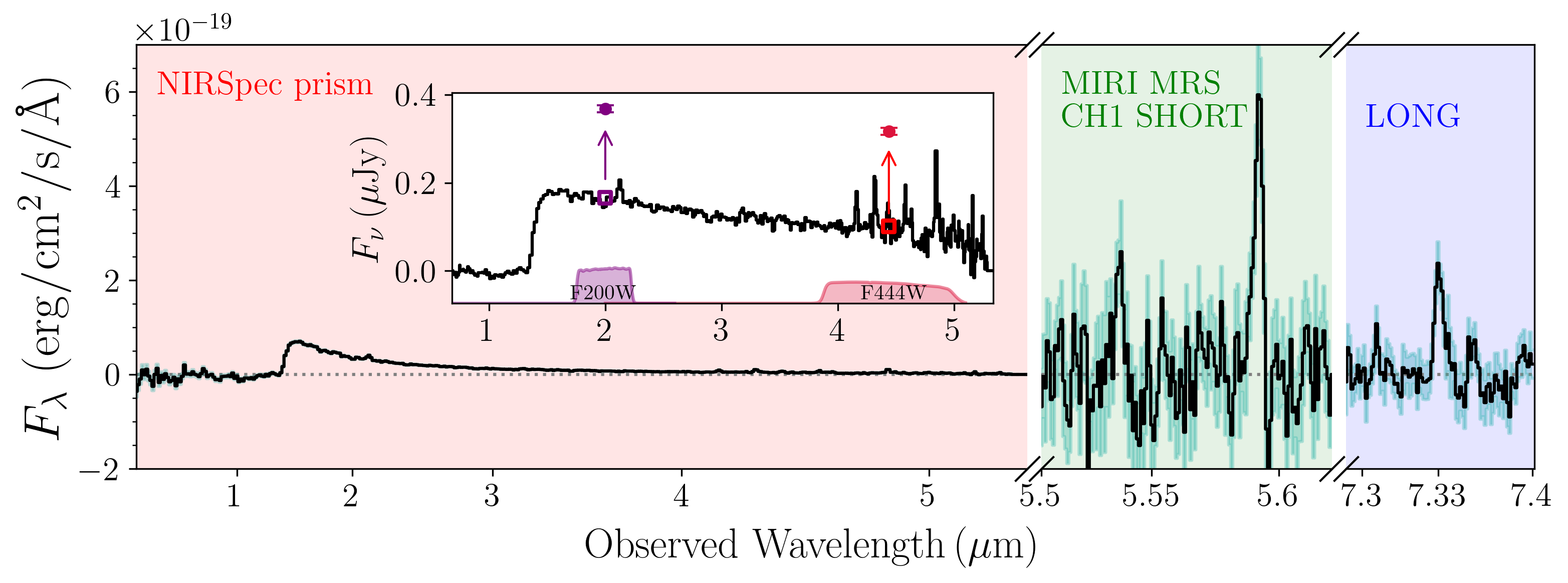}
\includegraphics[width=\textwidth]{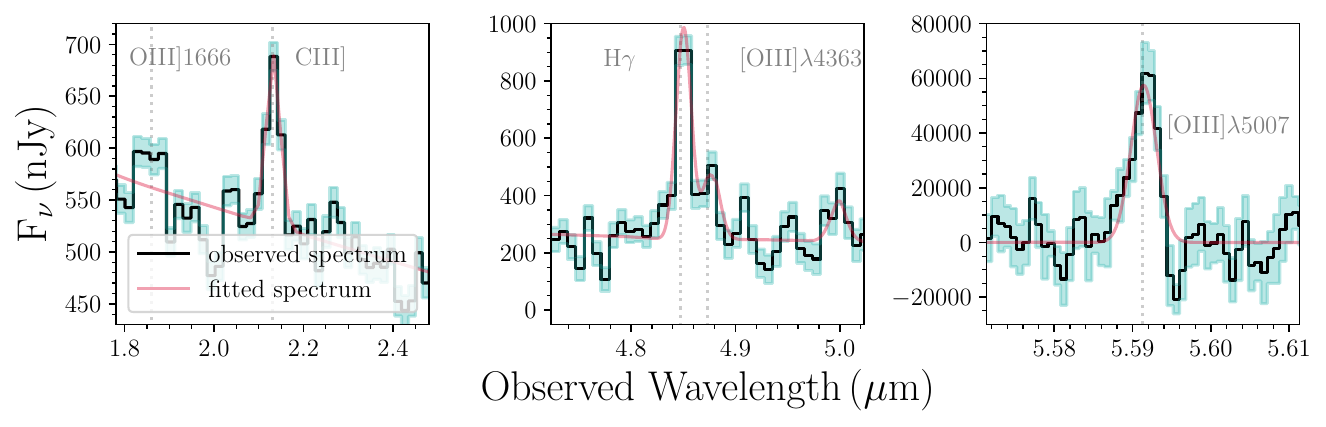}
\caption{{
(Top:)
The NIRSpec and MIRI spectrum of MACS0647$-$JD.
The inset shows how we normalize the NIRSpec data to NIRCam photometry with a JD1 magnification of $\mu=8$.
(Bottom:)
Emission lines used in this paper, including \CIIIdw\ (top panel), \OIIIwa\ (middle panel), and \OIIIw\ (bottom panel).
\CIIIdw\ and \OIIIwa\ shown are stacked NIRSpec spectra with total magnification $\mu=26.6$, while \OIIIw\ is the MIRI observation of JD1 with $\mu=8$.
The black lines show the observed spectrum and the red lines indicate the fit to the spectra and emission lines.
}
}
\label{fig:spec}
\end{figure*}
\input{table}


\section{Data and Measurement}
\label{Sec:data}
This article makes use of the NIRCam \citep{Rieke2005,Rieke2023} and NIRSpec \citep{NIRSpec_Jakobsen2022,NIRSpec-MOS_Ferruit2022,Boker2023} observations of JWST Cycle 1 program GO 1433 (PI Coe) and MIRI \citep{Rieke2015,Wright2015,Wright2023} observations of JWST Cycle 2 program GO 4246 (PI Abdurro'uf).
Both GO 1433 and GO 4246 observed MACS0647$-$JD.
The data are publicly available on MAST.\footnote{\url{https://mast.stsci.edu/search/ui/\#/jwst}\\ 
\dataset[DOI:10.17909/wpys-ap03]{https://archive.stsci.edu/doi/resolve/resolve.html?doi=10.17909/wpys-ap03},
\dataset[DOI:10.17909/re1k-jt10]{https://archive.stsci.edu/doi/resolve/resolve.html?doi=10.17909/re1k-jt10}
}

\subsection{NIRCam, NIRSpec, and MIRI}
\label{sec:jwst}

NIRCam imaging was obtained in 7 filters, including 6 wide-band filters, F115W, F150W, F200W, F277W, F356W, and F444W, and a medium band, F480M, spanning 1--5$\,\mu$m.
Exposure times were 2104 s in each filter and twice that in F200W.
All NIRCam data were reduced using the STScI JWST pipeline\footnoteurl{https://github.com/spacetelescope/jwst} \citep{JWST_pipeline_1.9.4} and \grizli\ \citep{grizli}.
In short, the pipeline performs corrections for $1/f$ noise striping
and masks ``snowballs''\footnote{\url{https://jwst-docs.stsci.edu/data-artifacts-and-features/snowballs-and-shower-artifacts}} 
and ``wisps''\footnote{\url{https://jwst-docs.stsci.edu/jwst-near-infrared-camera/nircam-instrument-features-and-caveats/nircam-claws-and-wisps}} in each NIRCam exposure and then drizzle-combines all exposures to a common $0\farcs02$ pixel grid.
Details on the NIRCam observations, data reduction, and photometric analysis of 
MACS0647$-$JD, including properties of the two individual components JDA and JDB, 
may be found in \citet{Hsiao2023a}.
Updated photometry was presented in \citet{Hsiao2024a}.
In this article, we use the measured flux densities of F200W $368\,{\rm nJy}$ and F444W $317\,{\rm nJy}$ for MACS0647$-$JD1 to correct the flux losses between NIRSpec MSA prism spectroscopy and MIRI IFU spectroscopy (see \S\ref{sec:normalization}).

NIRSpec multi-object spectroscopy (MOS) was performed using the microshutter assembly \citep[MSA;][]{Kutyrev2008,Rawle2022} to observe MACS0647$-$JD 
with the low-resolution prism ($R \sim 30-300$; 0.6 -- 5.3$\,$\um) 
with 3.6 hours exposure time split between two visits \citep{Hsiao2024a}.
Obs 23 performed standard 3-slitlet nods, 
while Obs 21 obtained data in single slilets with two dithers. 
Briefly, NIRSpec Level 1 data products were retrieved from MAST and were processed with the STScI JWST pipeline version 1.9.2
and \msaexp\footnoteurl{https://github.com/gbrammer/msaexp} version 0.6.0, to correct for 1/$f$ noise and mask snowballs.
For the single-slitlet data, we subtract 2D background spectra from a nearby slit that observed a relatively blank region of the image.
Full details of NIRSpec data reduction and background subtraction can be found in \citet{Hsiao2024a}.

In the final reduced and stacked spectrum of MACS0647$-$JD,
seven emission features were detected, including \CIIIdw, \OIIw, \NeIIIw, \NeIIIwb, \Hdeltaw, \Hgammaw, and the auroral line \OIIIwa\ \citep{Hsiao2024a}, {which is shown in Figure \ref{fig:spec}}.
In this article, we use emission line fluxes of the unresolved doublet \CIIIdw\ of ($428^{+34}_{-35}$)\e{-20} \cgsfluxunits\ and \OIIIwa\ of ($62 \pm 5$)\e{-20} \cgsfluxunits.
{We note that these measurements were} measured from the stacked spectrum of four prism observations (including two on JD1 ($\mu_{\rm JD1}=8$) and two on JD2 ($\mu_{\rm JD2}=5.3$)), with a total magnification of $\mu = 26.6$ \citep[][]{Hsiao2024a,Abdurrouf2024}.
{In Table \ref{tab:tab}, however, the fluxes are corrected for flux losses (see \S\ref{sec:normalization})}.
{Each line is fitted with a single Gaussian after continuum subtraction, and uncertainties are estimated using a Monte Carlo method.}
We show the emission lines in Figure \ref{fig:spec}.
Note that \citet{Hsiao2024a} did not detect \OIIIwb, which is the auroral line usually used to derive C/O.
The small bump seen at this wavelength is within the noise, while we also speculate the bump can include contributions from both \OIIIwb\ and \HeIIw, which are blended in the prism data.
Future NIRSpec grating spectroscopy could detect and resolve these features.

MIRI Medium Resolution Spectrograph (MRS) \citep{Wells2015,Argyriou2023} observed MACS0647$-$JD1 using integral field units (IFU) spectroscopy, covering all A, B and C components. 
The observations were conducted with two MRS bands, including SHORT and LONG, spanning 4.90--5.74\,$\mu$m and 6.53--7.65\,$\mu$m for channel 1, respectively.
Exposure times were 4.2 hours in each band.
MIRI spectroscopic data are processed with JWST pipeline version {1.13.4} and context {1215} of the Calibration Reference Data System (CRDS).
Details of the MIRI data reduction can be found in \citet{Hsiao2024b}.
{We also present the MIRI spectroscopic data in Figure \ref{fig:spec}.}
\OIIIww\ was detected in the SHORT band with a line flux of ($226 \pm 21$)\e{-19} \cgsfluxunits, and \Halpha\ was detected in the LONG band with a line flux of ($90 \pm 10$)\e{-19} \cgsfluxunits\ \citep{Hsiao2024b}.
Note that \Hbeta\ was not detected given the short exposure time.
{
Emission lines detected with MIRI are fitted with Gaussian functions, along with an additional second-order polynomial to model the local background, differing from the fitting procedure used for NIRSpec emission lines.
Uncertainties are estimated using a Monte Carlo approach, based on 1000 noise realizations generated from the measured continuum RMS.
Based on the reported electron temperature $T_{\rm e}$(\OIII) $= 17000 \pm 1000\,$K  and electron density log($n_{e}$)$=2.9\pm0.5$ \citep{Abdurrouf2024,Hsiao2024b}, , the theoretical Balmer line ratio  H$\alpha$/H$\gamma$ is approximately 5.84.
Given that our measured value is H$\alpha$/H$\gamma$ = 5.5$\pm$0.7, which aligns with the theoretical expectation within uncertainties, we assume negligible dust attenuation throughout this analysis.}

\subsection{NIRSpec spectroscopy normalized to MIRI MRS}
\label{sec:normalization}

In the MIRI MRS, JD1 A+B are both covered by the IFU, and the line flux measurements are integrated over a 0\farcs25 aperture, with a correction for the aperture losses \citep[see \S 3.1.1 in][]{Hsiao2024b}.
However, for the slitlet spectroscopy NIRSpec MSA, the slits did not fully cover JD1 A+B.
In order to account for the line ratios between MIRI (\OIIIw) and NIRSpec (\CIIIdw\ and \OIIIwa), we follow a similar approach as in \citet{Hsiao2024b} to correct for the flux losses. 

\cite{Hsiao2024a} measured NIRCam photometry of MACS0647$-$JD within apertures of $r = 0.25\arcsec$, including aperture corrections.
We integrate the NIRSpec stacked spectra over both F200W and F444W filters, which are the filters that cover \CIIIdw\ and \OIIIwa, respectively.
We measure 189$\,$nJy and 105$\,$nJy for the stacked spectrum in the bandpass of F200W and F444W (with JD1 magnification $\mu = 8$), respectively.
The different correction factors for \CIIIdw\ and \OIIIwa\ might be due to the blue and red nature of the spectrum for JDA and JDB.
Then the correction between the fluxes integrated with the NIRSpec prism is multiplied to match the NIRCam aperture photometry to correct for slit losses and apply other factors as needed to correct for magnification {(i.e., the fluxes quoted in \citet{Abdurrouf2024} were measured from the stacked spectrum with a total magnification of $\mu=26.6$ and we scale them to the JD1 magnification of $\mu=8$, as shown in Table \ref{tab:tab})}.
{This process is visualized in the inset of Figure \ref{fig:spec}.}

We also estimate line ratios of \CIIIdw\ $/$ \OIIIw\ $=0.11\pm0.01$ and \OIIIw\ $/ \lambda4363 = 40\pm5$.
The measurements and the properties used are organized in Table \ref{tab:tab}.

\begin{figure*}
\centering
\includegraphics[width=\textwidth]{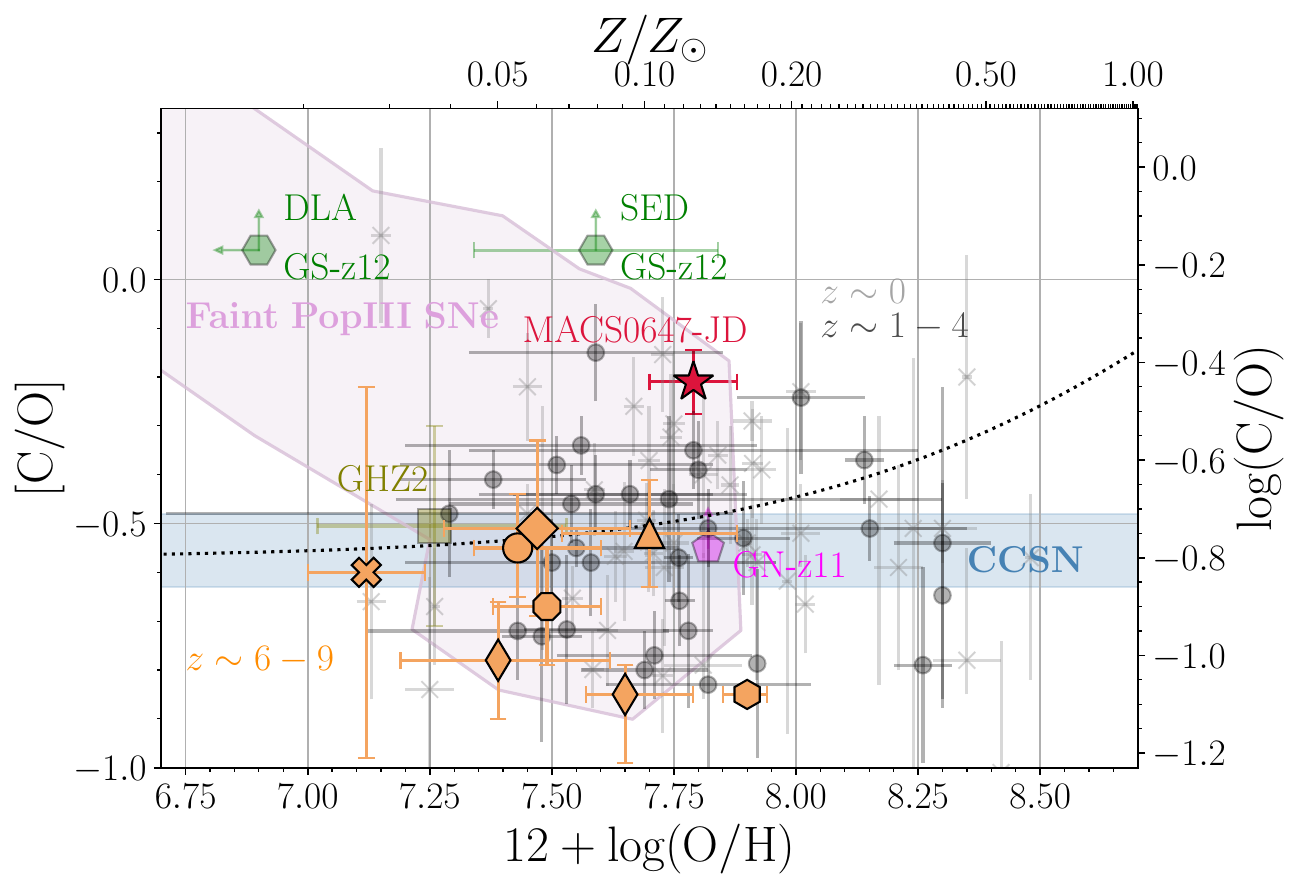}
\caption{
Abundance ratios [C/O] (relative to solar) and \logOH\ for MACS0647$-$JD compared to other galaxies.
{We include low-z galaxies \citep[the grey crosses;][]{Berg2016,Senchyna2017,Berg2019}, Milky Way Stars \citep{Nicholls2017}, {intermediate-z ($z\sim1-4$) galaxies \citep[the black circles; e.g.,][]{Pettini2000,Fosbury2003,Erb2010,Rigby2011,Christensen2012,Bayliss2014,James2014,Stark2014,Steidel2016,Vanzella2016,Amorin2017,Rigby2018,Berg2018,Citro2024}}, 
as well as high-z galaxies recently studied spectroscopically with JWST: 
GS-z12 \citep[$z=12.5$; the green hexagon;][]{DEugenio2024}, GN-z11 \citep[$z=10.6$; the magenta petagon;][]{Cameron2023}, GHZ2 \citep[$z=12.34$; the olive square;][]{Castellano2024}, RXCJ2248$-$ID \citep[$z=6.11$; the orange circle;][]{Topping2024}, GLASS$-$150008 \citep[$z=6.23$; the orange diamonds;][]{Jones2023,Isobe2023}, MACS1149$-$JD1 \citep[$z=9.11$; the orange hexagon;][]{Stiavelli2023},
GS$-$z9$-$0 \citep[$z=9.43$; the orange octogon;][]{Curti2024}, CEERS$-$1019 \citep[$z=8.68$; the orange triangle;][]{Marques2024}, A1703$-$zd6 \citep[$z=7.04$; the (fat) orange diamond;][]{Topping2025}, and s04590 \citep[$z=8.495$; the orange cross;][]{Arellano2022}.
Note that galaxies at $z>10$ in this plot, except MACS0647$-$JD, are derived either assuming electron temperature \citep{Cameron2023,Castellano2024} or through SED modelling \citep{DEugenio2024}, displayed with partial transparency.
For GS-z12, \citet{DEugenio2024} presented DLA metallicity for 12 + log(O/H); we also show their 12 + log(O/H) from their \texttt{BEAGLE} {SED fitting of $7.59\pm0.25$}.
{We also compare the theoretical predictions of the relation, including metal-poor core-collapse supernovae  \citep[CCSN;][]{Tominaga2007,Curti2025}, shown in the {blue} region, and  {faint supernovae from pure Pop {\sc iii}} \citep[][]{Vanni2023,Curti2025}, shown in the {magenta} region.}
}}
\label{fig:co}
\end{figure*}

\section{Method and Result}
\label{Sec:method}

{In order to obtain C/O, we first calculate higher excitation $\rm{C^{2+}}/\rm{O^{2+}}$, and correct for the unobserved ionic species with an ionization correction factor} \citep[ICF; e.g.,][]{Berg2019,Castellano2024}:
\begin{equation}
    \rm{\frac{C}{O}}=\frac{\rm{C^{2+}}}{\rm{O^{2+}}} \times {\rm ICF}.
\end{equation}
With the detected line fluxes of \OIIIw\ from MIRI MRS and the auroral line \OIIIwa\ from NIRSpec, we obtain an electron temperature of $T_{\rm e}$(\OIII) $= 17000 \pm 1000\,$K
using the \pyneb\ \citep{pyneb_Luridiana2015} task \texttt{getTemDen} \citep{Hsiao2024b}.
We then assume that the regions emitting \CIIIdw\ have the same electron temperature as those emitting \OIIIw.

For the electron density, we adopt log($n_{e}$)$=2.9\pm0.5$ derived by \citet{Abdurrouf2024}
based on resolved \OIIdw\ assuming a uniform density model.
(See \S\ref{Sec:future} for future prospects on measuring electron densities in higher-ionization zones from different lines.)
Note that the electron density does not significantly affect our result,
with log(C/O) changing by $< 0.01$ when varying log($n_{e}$) {between $2.4$ to $3.4$.}
As a result, ionic abundances of $\rm{O^{2+}}/\rm{H^{2+}}$ and $\rm{C^{2+}}/\rm{H^{2+}}$ can be obtained using the task \texttt{getIonAbundance}. 

For the ICF, we follow the calibration from \citet{Berg2019}.
With a ${\rm log}(U)=-1.9$ from \OIIIw/\OIIw\ and metallicity $Z=0.1\,Z_{\odot}$ \citep{Hsiao2024b}, we estimate an ICF of $1.10\pm0.04$.
Therefore, after the ICF correction, we obtain $\rm{log(C/O)}=-0.44^{+0.06}_{-0.07}$.
These measurements are presented in Table \ref{tab:tab}.


\section{Discussion}
\label{Sec:discussion}

\subsection{A high C/O in the first 500$\,$Myr of cosmic time?}
\label{Sec:highco}

We find a sub-solar relative carbon abundance in MACS0647$-$JD.
Figure \ref{fig:co} shows the relation between 
carbon abundance relative to solar [C/O]
and metallicity $12+{\rm log(O/H)}$.
{We compare MACS0647$-$JD with previous low-z measurements \citep[$z\sim0$; e.g.,][]{Berg2016,Senchyna2017,Berg2019}, {intermediate-z \citep[$z\sim1-4$; e.g.,][]{Pettini2000,Fosbury2003,Erb2010,Rigby2011,Christensen2012,Bayliss2014,James2014,Stark2014,Steidel2016,Vanzella2016,Amorin2017,Rigby2018,Berg2018,Citro2024}} and high-z galaxies recently observed with \JWST.}
More metal-rich galaxies generally have higher C/O ratios, as in the case with stellar abundances in the Milky Way \citep{Nicholls2017}, shown in the black-dotted line.

Some recent galaxies at $6<z<10$ observed with JWST also present a similar trend {of low C/O ratios at low O/H}, 
{including s04590 at $z=8.495$ \citep{Arellano2022}, GS-z9-0 at $z=9.43$ \citep{Curti2024}, CEERS-1019 at $z=8.68$ \citep{Marques2024},
RXCJ2248$-$ID at $z=6.11$ \citep{Topping2024}, A1703-zd6 at $z=7.04$ \citep{Topping2025}
and even lower C/O galaxies such as 
GLASS$-$150008 at $z=6.23$ \citep{Jones2023,Isobe2023} and MACS1149$-$JD1 at $z=9.11$ \citep{Stiavelli2023}.}
All of these galaxies are metal-poor ($Z<0.2\,Z_{\odot}$) and C/O-poor (log(C/O)$<-0.75$) at $6<z<10$, with direct $T_{e}$ measurements.

However, in the $z>10$ regime, \OIIIwa\ and \OIIIw\ are difficult to detect due to faintness and redshift, respectively.
With assumed $T_{e}$, \cite{Castellano2024} estimate GHZ2 ($z=12.34$) having a C/O as low as the $6 < z < 10$ galaxies mentioned above.
For GN-z11 ($z=10.6$), \citet{Cameron2023} estimated a lower limit suggesting higher C/O.
However, GS-z12 ($z=12.5$) hosts a significantly super-solar carbon abundance \citep{DEugenio2024}.
They interpreted that such a high C/O ratio can be explained by the {low-energy supernovae ($<2\times10^{51}\,{\rm erg}$) from Pop {\sc iii} stars} given the extremely {low metallicity of 12+log(O/H)$<6.9$} \citep[see also][]{Maiolino2019,Vanni2023}.

A caveat of the $z>10$ results is that none of those galaxies have direct metallicity or direct C/O measurements.
Especially in GS-z12, the lack of Balmer lines and \OIIIw\ makes the O/H and C/O uncertain.
\citet{DEugenio2024} estimated C/O based on a strong CIII] detection ($30 \pm 7$ \AA\ EW)
and upper limits on \OIIIwb, \OII, and [NeIII].
The lines \OIIIwa\ and \OIIIw, plus most of the Balmer lines (\Halpha, \Hbeta, \Hgamma, \Hdelta),
are all redshifted beyond NIRSpec's coverage, 
and they lack MIRI data.
They estimated the gas metallicity O/H two ways: 
using the properties of the local DLA (column density and dust reddening) 
and from \texttt{BEAGLE} SED fitting;
we include both measurements in Fig.~\ref{fig:co}.

In this work, MACS0647$-$JD shows a slightly higher C/O compared with most other high-z galaxies observed with \JWST, probably due to its higher metallicity of 12+log(O/H)$=7.79\pm0.09$.
We note the C/O in MACS0647$-$JD is higher than the local trend found in \citet{Nicholls2017}, but may be within the scatter.
We emphasize that our measurement marks a milestone and the first-ever direct C/O at $z>10$, 
including direct metallicity via $T_{e}$ and electron density $n_{e}$,
which is the most precise way to estimate the chemical abundance. 

MACS0647$-$JD has a mass-weighted stellar age of approximately $50\,$Myr \citep{Hsiao2023a}.
If long-lived intermediate-mass stars are the dominant source of carbon, such a high C/O ratio would not be expected at this early stage. 
However, this enrichment may be explained by a quiescent phase followed by a recent burst of star formation.
With a formation age of $\sim150\,$Myr \citep{Hsiao2023a}, the first AGB stars could have started releasing carbon into the ISM as early as $\sim50\,$Myr after the onset of star formation, though the majority of carbon is typically produced over timescales of several hundred Myr.
The observed elevated C/O ratio, therefore, suggests early metal enrichment, occurring just $\sim400\,$Myr after the Big Bang.

There are also a few local galaxies with similar C/O and O/H studied by \citet{Berg2019}, suggesting that MACS0647$-$JD may have undergone a very efficient and rapid burst of star formation, or has a low effective oxygen abundance yield ({more oxygen outflow;} see \citealt{Berg2019} Figure 12 therein) and could hint that very bursty star formation is important for star formation in the first few million years.
Such higher C/O ratio can also originate from exotic stellar populations, including the super massive stars in proto globular clusters \citep{Charbonnel2023} or possibly Pop {\sc iii} heritage if extremely low metallicity \citep{Vanni2023,DEugenio2024}.
{As shown in Figure \ref{fig:co}, the C/O versus O/H relation in MACS0647$-$JD cannot be explained by Population II core-collapse supernovae (CCSNe), but is consistent with enrichment from {faint (low-energy) supernovae from pure Pop {\sc iii}}.}

In the {interpretation above}, we ignore the fact that MACS0647$-$JD consists of two components, JDA and JDB.
{
JDA has a relatively young formation age of only a few Myr, whereas JDB is older, with an estimated age of $\sim100-200\,$Myr, according to the delayed$-\tau$ model \citep[Figure 13;][]{Hsiao2023a}.
As mentioned earlier, MACS0647$-$JD may have hosted the first AGB stars formed $\sim150\,$Myr before the observation.
These AGB stars may have originated from JDB, leading to a higher C/O ratio when observed.
This is consistent with the earlier hypothesis that MACS0647$-$JD could have a young age overall while still hosting old enough stellar populations for AGB enrichment.}
The 2D prism spectra in \citet{Hsiao2024a} and \citet{Abdurrouf2024} {also} show JDB may contribute significantly to \CIIIdw, {further supporting this scenario.}
However, we note that it is rather difficult to model two components and disentangle them in the slit spectrum without integral field unit (IFU) spectroscopy (see \S\ref{Sec:future}), as we mentioned in \S\ref{sec:normalization}.
{Furthermore, we also note that the timescales for stellar mass build-up in JDA and JDB are only $\sim460\,$Myr, which might be shorter than the timescale required for low- to intermediate-mass stars to contribute significantly to carbon enrichment \citep[$\sim0.1-1$Gyr; see, e.g.,][]{Mattsson2010}.}




\subsection{Future Spectroscopic Observations}
\label{Sec:future}

As discussed in \S\ref{Sec:highco}, MACS0647$-$JD has two components A and B that are hard to disentangle as they were observed in different modes with NIRSpec MSA and MIRI MRS IFU.
Therefore, NIRSpec IFU spectroscopic observations are required to further identify the origins of \CIIIdw\ and other oxygen and hydrogen lines.
JWST Cycle 3 program GTO 4528 (PI Isaak) will observe MACS0647$-$JD with NIRSpec IFU G395M, which will detect rest-frame optical lines (2570 -- 4565\AA) in JDA and JDB individually.
It can also potentially detect \OIIIwa\ in JDA and JDB and provide the direct metallicity in each clump, combined with careful modeling of A+B in the MIRI IFU spectrum.
Future NIRSpec IFU observations with G235M/H (covering 1490 -- 2750\AA) are essential and required to study C/O in the two components separately.
G235H would resolve \CIIIdw, delivering a more precise density in that regime rather than assuming the density derived from \OIIdw\ \citep[e.g.,][]{Acharyya2019,Mingozzi2022,Topping2024}.
While log(C/O) $\sim -0.44$ only changes by 0.01 when varying $\rm{log}(n_e / \rm{cm^{-3}}) = 2.9 \pm 0.5$ within the uncertainties,
it could change more at significantly higher densities, 
for example log(C/O) = $-0.65$ for $\rm{log}(n_e / \rm{cm^{-3}}) = 5$.

Not only for MACS0647$-$JD, but also other $z>10$ galaxies mentioned are worth follow-up observations.
For instance, MIRI would detect \OIIIw\ in GS-z12, which will enable more precise C/O (and also O/H) measurements.
MIRI MRS observations have been obtained for GN-z11 (GO 2926; PI Colina).
Deeper NIRSpec spectroscopic observations on these galaxies could deliver auroral lines and Balmer lines, leading to more accurate ``direct" metallicity and C/O measurements.
More generally, more galaxies at $z>6$, or even $z>3$ with spectroscopic data, would help answer whether C/O has a similar redshift evolution between $0<z<3$ as in the mass-metallicity relation and help us understand how carbon is enriched through different processes in the early universe.

\section{Conclusions} \label{sec:conclusion}
In this article, we estimate the C/O abundance in a triply-lensed galaxy MACS0647$-$JD at $z=10.17$.
MACS0647$-$JD showed a bright \CIIIdw\ feature in the NIRSpec prism spectrum while no \OIIIwb\ is detected.
We estimate ${\rm log(C/O)}=-0.44^{+0.06}_{-0.07}$ using the direct electron temperature method for the first time at $z>10$.
MACS0647$-$JD is also the only $z>10$ galaxy with a direct metallicity measurement.
In low-z galaxies, C/O increases with metallicity 12+log(O/H).
MACS0647$-$JD has a higher C/O ratio than local galaxies of similar metallicity, possibly due to a very
efficient and rapid burst of star formation, a low effective oxygen abundance yield, or even exotic stellar populations.

We also suspect JDB, one of the component star clusters, may be the culprit of higher C/O.
JDB is older, as revealed from NIRCam photometry SED fitting, suggesting it might host relatively abundant carbon since the intermediate-mass stars have started injecting carbon into the ISM.
Future NIRSpec IFU observations, especially G235M/H, are essential to separate two components JDA and JDB.
Additional deep spectroscopic observations of other galaxies are required to determine statistically whether a higher carbon abundance trend exists in the early universe and possibly constrain when the first stars formed.

\section{Acknowledgments}
We appreciate Prof.~Daniel Eisenstein, Prof.~Dan Stark, Dr.~Mirko Curti, Dr.~Francesco D'Eugenio and the anonymous referee for insightful comments.

This work is based on observations made with the NASA/ESA/CSA 
\textit{James Webb Space Telescope} (JWST). 
The data were obtained from the Mikulski Archive for Space Telescopes (MAST) 
at the Space Telescope Science Institute (STScI), 
which is operated by the Association of Universities for Research in Astronomy (AURA), Inc., 
under NASA contract NAS 5-03127 for JWST. 
We are grateful and indebted to the 20,000 people who worked to make JWST an incredible discovery machine.
These observations are associated with JWST programs GO 4246 and 1433.
TH and A are funded by grants for JWST-GO-4246 provided by STScI under NASA contract NAS5-03127.
TH appreciates the support from the Government scholarship to study abroad (Taiwan).

PD acknowledge support from the NWO grant 016.VIDI.189.162 (``ODIN") and warmly thanks the European Commission's and University of Groningen's CO-FUND Rosalind Franklin program. 

J.A-M. acknowledge support by grants PIB2021-127718NB-100 from the Spanish Ministry of Science and Innovation/State Agency of Research MCIN/AEI/10.13039/501100011033 and by “ERDF A way of making Europe”.



%

\vspace{5mm}





\bibliography{papers}{}
\bibliographystyle{aasjournal}



\end{document}

%% file: newcommands.tex
\newcommand{\LCDM}{$\Lambda$CDM}

\newcommand{\red}[1]{{\color{red} #1}}
\newcommand{\redss}[1]{{\color{red} ** #1}}
\newcommand{\redbf}[1]{{\color{red}\bf #1 \color{black}}}

\newcommand{\ny}{$\tilde {\rm n}$}
\newcommand{\about}{$\sim$}
\newcommand{\appr}{$\approx$}
\newcommand{\gt}{$>$}
\newcommand{\um}{$\mu$m}
\newcommand{\uJy}{$\mu$Jy}
\newcommand{\sig}{$\sigma$}
\newcommand{\Lya}{Lyman-$\alpha$}
\renewcommand{\th}{$^{\rm th}$}
\newcommand{\lam}{$\lambda$}

\newcommand{\tentothe}[1]{$10^{#1}$}
\newcommand{\tentotheminus}[1]{$10^{-#1}$}
\newcommand{\e}[1]{$\times 10^{#1}$}
\newcommand{\en}[1]{$\times 10^{-#1}$}
\newcommand{\cgsfluxunits}{erg$\,$s$^{-1}\,$cm$^{-2}$}
\newcommand{\linefluxunits}{\tentotheminus{20} \cgsfluxunits}

\newcommand{\logU}{$\log(U)$}
\newcommand{\logOH}{12+log(O/H)}

\newcommand{\sinv}{s$^{-1}$}
\newcommand{\kms}{km\,s$^{-1}$}

\newcommand{\footnoteurl}[1]{\footnote{\url{#1}}}

\newcommand{\tnm}[1]{\tablenotemark{#1}}
\newcommand{\super}[1]{$^{\rm #1}$}
\newcommand{\supa}{$^{\rm a}$}
\newcommand{\supb}{$^{\rm b}$}
\newcommand{\supc}{$^{\rm c}$}
\newcommand{\supd}{$^{\rm d}$}
\newcommand{\supe}{$^{\rm e}$}
\newcommand{\supf}{$^{\rm f}$}
\newcommand{\supg}{$^{\rm g}$}
\newcommand{\suph}{$^{\rm h}$}
\newcommand{\supi}{$^{\rm i}$}
\newcommand{\supj}{$^{\rm j}$}
\newcommand{\supk}{$^{\rm k}$}
\newcommand{\supl}{$^{\rm l}$}
\newcommand{\supm}{$^{\rm m}$}
\newcommand{\supn}{$^{\rm n}$}
\newcommand{\supo}{$^{\rm o}$}

\newcommand{\squared}{$^2$}
\newcommand{\cubed}{$^3$}

\newcommand{\sqarcmin}{arcmin\squared}

\newcommand{\supcomma}{$^{\rm ,}$}

\newcommand{\rhalf}{$r_{1/2}$}

\newcommand{\chisq}{$\chi^2$}

\newcommand{\Zgas}{$Z_{\rm gas}$}  
\newcommand{\Zstar}{$Z_*$}  

\newcommand{\per}{$^{-1}$}
\newcommand{\inv}{\per}
\newcommand{\Mstar}{$M^*$}
\newcommand{\Lstar}{$L^*$}
\newcommand{\phistar}{$\phi^*$}

\newcommand{\logM}{log($M_*$/\Msun)}

\newcommand{\LUV}{$L_{UV}$}
\newcommand{\MUV}{$M_{UV}$}

\newcommand{\Msun}{$M_\odot$}
\newcommand{\Lsun}{$L_\odot$}
\newcommand{\Zsun}{$Z_\odot$}

\newcommand{\Mvir}{$M_{vir}$}
\newcommand{\Mt}{$M_{200}$}
\newcommand{\Mf}{$M_{500}$}

\newcommand{\Ndotion}{$\dot{N}_{\rm ion}$}
\newcommand{\xiion}{$\xi_{\rm ion}$}
\newcommand{\logxiion}{log(\xiion)}
\newcommand{\fesc}{$f_{\rm esc}$}

\newcommand{\XHI}{$X_{\rm HI}$}
\newcommand{\XHII}{$X_{\rm HII}$}
\newcommand{\RHII}{$R_{\rm HII}$}

\newcommand{\Halpha}{H$\alpha$}
\newcommand{\Hbeta}{H$\beta$}
\newcommand{\Hgamma}{H$\gamma$}
\newcommand{\Hdelta}{H$\delta$}
\newcommand{\Halphaw}{\Halpha\,$\lambda$6563}
\newcommand{\Hbetaw}{\Hbeta\,$\lambda$4861}
\newcommand{\Hgammaw}{H$\gamma$\,$\lambda$4340}
\newcommand{\Hdeltaw}{H$\delta$\,$\lambda$4101}
\newcommand{\Ha}{\Halpha}
\newcommand{\Hb}{\Hbeta}

\newcommand{\I}{\,{\sc i}}
\newcommand{\II}{\,{\sc ii}}
\newcommand{\III}{\,{\sc iii}}
\newcommand{\IV}{\,{\sc iv}}
\newcommand{\V}{\,{\sc v}}
\newcommand{\VI}{\,{\sc vi}}
\newcommand{\VII}{\,{\sc vii}}
\newcommand{\VIII}{\,{\sc viii}}

\newcommand{\HI}{H\I}
\newcommand{\HII}{H\II}
\newcommand{\HeI}{He\I}
\newcommand{\HeII}{He\II}

\newcommand{\CII}{[C\II]}
\newcommand{\CIIw}{\CII\,$\lambda$2325 (blend)}
\newcommand{\CIII}{[C\III]}
\newcommand{\CIIIw}{\CIII\,$\lambda$1908}
\newcommand{\CIIId}{C\III]}
\newcommand{\CIIIdw}{C\III]\,$\lambda\lambda$1907,1909}
\newcommand{\CIV}{C\IV}
\newcommand{\CIVw}{\CIV\,$\lambda$1549}
\newcommand{\OII}{[O\II]}
\newcommand{\OIIw}{\OII\,$\lambda$3727}
\newcommand{\OIIdw}{\OII\,$\lambda\lambda$3727,3729}
\newcommand{\OIII}{[O\III]}
\newcommand{\OIIIw}{\OIII\,$\lambda$5008}
\newcommand{\OIIIww}{\OIII\,$\lambda$4960,$\lambda$5008}
\newcommand{\OIIIdw}{\OIIIww}
\newcommand{\OIIIwa}{\OIII\,$\lambda$4363}
\newcommand{\OIIIwb}{O\III]\,$\lambda$1666}
\newcommand{\OIIIwc}{\OIII\,$\lambda$4960}
\newcommand{\NeIII}{[Ne\III]}
\newcommand{\NeIIIw}{\NeIII\,$\lambda$3869}
\newcommand{\NeIIIwb}{\NeIII\,$\lambda$3968}
\newcommand{\NII}{[N\II]}
\newcommand{\NIIw}{\NII\,$\lambda$6585}
\newcommand{\NIIww}{\NII\,$\lambda$6550,$\lambda$6585}
\newcommand{\SII}{[S\II]}
\newcommand{\SIIww}{\SII\,$\lambda$6718,$\lambda$6733}
\newcommand{\HeIw}{HeI\,$\lambda$3889}
\newcommand{\HeIwa}{HeI\,$\lambda$4473}
\newcommand{\HeIIw}{HeII\,$\lambda$1640}
\newcommand{\NIII}{N\III]}
\newcommand{\NIV}{N\IV]}
\newcommand{\NIIIw}{\NIII\,$\lambda$1748}
\newcommand{\NIVw}{\NIV\,$\lambda$1486}
\newcommand{\MgII}{Mg\II}
\newcommand{\MgIIw}{\MgII\,$\lambda$2800}

\newcommand{\Lyaw}{Ly$\alpha$\,$\lambda$1216}



\newcommand{\Om}{\Omega_{\rm M}}
\newcommand{\OL}{\Omega_\Lambda}

\newcommand{\etal}{et al.}

\newcommand{\citeps}{\citep}

\newcommand{\HST}{{\em HST}}
\newcommand{\SST}{{\em SST}}
\newcommand{\Hubble}{{\em Hubble}}
\newcommand{\Spitzer}{{\em Spitzer}}
\newcommand{\Chandra}{{\em Chandra}}
\newcommand{\JWST}{{\em JWST}}
\newcommand{\Planck}{{\em Planck}}

\newcommand{\Bradac}{{Brada\v{c}}}

\newcommand{\citepeg}[1]{\citep[e.g.,][]{#1}}

\newcommand{\range}[2]{\! \left[ _{#1} ^{#2} \right] \!}  

\newcommand{\grizli}{\textsc{grizli}}
\newcommand{\eazypy}{\textsc{eazypy}}
\newcommand{\msaexp}{\textsc{msaexp}}
\newcommand{\trilogy}{\textsc{trilogy}}
\newcommand{\bagpipes}{\textsc{bagpipes}}
\newcommand{\beagle}{\textsc{beagle}}
\newcommand{\photutils}{\textsc{photutils}}
\newcommand{\SEP}{\textsc{sep}}
\newcommand{\piXedfit}{\textsc{piXedfit}}
\newcommand{\pyneb}{\textsc{pyneb}}
\newcommand{\HIIC}{\textsc{hii-chi-mistry}}
\newcommand{\astropy}{\textsc{astropy}}
\newcommand{\astrodrizzle}{\textsc{astrodrizzle}}
\newcommand{\multinest}{\textsc{multinest}}
\newcommand{\cloudy}{\textsc{Cloudy}}
\newcommand{\jdaviz}{\textsc{Jdaviz}}

\renewcommand{\tt}[1]{\texttt{#1}}

\newcommand{\SE}{\tt{SourceExtractor}}

\newcommand{\PD}[1]{\textcolor{blue}{[PD: #1\;]}}

%% file: table.tex
\begin{deluxetable}{cc}
\tablecaption{\label{tab:tab}Emission lines, lines ratios, and physical properties estimated for MACS0647$-$JD.}
\tablewidth{\columnwidth}
\tablehead{
\colhead{} &
\colhead{}
}
\startdata
\multicolumn{2}{c}{Emission Line Fluxes ($10^{-19}\,$erg/s/cm$^2$)} \\ \hline
\CIIIdw$^{a}$ & $25\pm2$ \\
\OIIIwa$^{a}$ & $5.6\pm0.5$ \\
\OIIIw$^{b}$ & $226\pm21$ \\ \hline
\multicolumn{2}{c}{Emission Line Ratios} \\ \hline
\OIIIw\ $/$ $\lambda$4363$^{b}$ & $40\pm5$  \\
O32$^{b}$ = \OIIIw\ $/$ \OIIw\ & $17\pm2$ \\
H$\alpha$ / H$\gamma^{b}$ & $5.5\pm0.7$  \\
R3$^{b}$ = \OIIIw\ $/$ \Hbeta & {$6.9\pm1.0$}\ \\
\CIIIdw\ $/$ \OIIIw\ & {$0.11\pm0.01$}\ \\
\hline \hline   
\multicolumn{2}{c}{Physical Properties} \\ \hline  
12+log(O/H)$^{b}$ & $7.79\pm0.09$ \\ 
log($U$)$^{b}$ & $-1.9\pm0.1$ \\
$T_{e}(\rm{[OIII]})$(K)$^{b}$ & $17000\pm1000$ \\
log($n_{e}$)$^{c}$ & $2.9\pm0.5$ \\
C ICF & $1.10\pm0.04$ \\
${\rm C^{2+}/O^{2+}}$ & $0.33\pm0.05$\\
log(C/O) & $-0.44^{+0.06}_{-0.07}$
\enddata
\tablenotetext{a}{Corrected for flux losses to match MIRI IFU.}
\tablenotetext{b}{\citet{Hsiao2024b}}
\tablenotetext{c}{\citet{Abdurrouf2024}.}
\end{deluxetable}